# Raman spectroscopy of epitaxial graphene on a SiC substrate

Z. H. Ni,[1] W. Chen,[2] X. F. Fan,[1] J. L. Kuo,[1] T. Yu,[1]* A. T. S. Wee,[2] Z. X. Shen[1]*

[1] Division of Physics and Applied Physics, School of Physical & Mathematical Sciences, Nanyang Technological University, Singapore 637371
[2] Department of Physics, Faculty of Science, 2 Science Drive 3, National University of Singapore, Singapore 117542

**Abstract**

The fabrication of epitaxial graphene (EG) on SiC substrate by annealing has attracted a lot of interest as it may speed up the application of graphene for future electronic devices. The interaction of EG and the SiC substrate is critical to its electronic and physical properties. In this work, Raman spectroscopy was used to study the structure of EG and its interaction with SiC substrate. All the Raman bands of EG blue shift from that of bulk graphite and graphene made by micromechanical cleavage, which was attributed to the compressive strain induced by the substrate. A model containing 13 × 13 honeycomb lattice cells of graphene on carbon nanomesh was constructed to explain the origin of strain. The lattice mismatch between graphene layer and substrate causes the compressive stress of 2.27 GPa on graphene. We also demonstrate that the electronic structures of EG grown on Si and C terminated SiC substrates are quite different. Our experimental results shed light on the interaction between graphene and SiC substrate that are critical to the future applications of EG.

PACS: 61.46.-w, 78.30.−j, 68.35.Gy

Corresponding authors: yuting@ntu.edu.sg ; zexiang@ntu.edu.sg

**Introduction**

Graphene comprises one monolayer of carbon atoms packed into a two-dimensional (2D) honeycomb lattice.[1] It has attracted much interest since it was firstly discovered in 2004.[2,3] The electrons in an ideal graphene sheet behave like massless Dirac-Fermions.[4,5] Therefore, graphene exhibits a series of new electronic properties such as anomalously quantized Hall effects, absence of weak localization and the existence of a minimum conductivity.[1-3] The peculiar properties of graphene make it a promising candidate for fundamental studies as well as for potential device applications.[6-10]

Two approaches have been successfully developed for fabrication of graphene: micromechanical cleavage of graphite [2,3] and epitaxial growth on silicon carbide (SiC) substrate.[11,12] The former can be used to obtain high quality graphene sheets which are comparable to that in graphite, but is restricted by small sample dimensions and low visibility. Epitaxial graphene (EG) grown on SiC is suitable for large area fabrication is more compatible with current Si processing techniques for future applications. Nevertheless, the EG may interact with the SiC substrate which could modify its optical and electronic properties. A bandgap of ~0.26 eV was observed by angle-resolved photoemission spectroscopy (ARPES) on EG grown on SiC substrate, which attributed to the interaction of graphene with the substrate.[13] Some theoretical[14,15] and experimental studies on EG, e.g. X-ray diffraction (XRD)[15,16] and scanning tunneling microscopy (STM)[17,18], have also been carried out. However the effect of SiC substrate on EG is still not well understood. In previous studies,[14-19] the formation of graphene on SiC substrate can be described as follows: the SiC surface first reconstructs to a ($\sqrt{3}\times\sqrt{3}$)R30° (R3) structure, then to a ($6\sqrt{3}\times6\sqrt{3}$)R30° (6R3) structure, referred as carbon nanomesh in this paper; after higher temperature annealing, the graphene/graphite forms on carbon nanomesh. However, it is still under debate as to how the graphene bonds/connects to the 6R3/carbon nanomesh structure.

Raman spectroscopy has been extensively used in the study of graphene. For example, the second order (2D) Raman band is used as a simple and efficient way to identify the single layer graphene made by micromechanical cleavage;[19,20] Raman spectroscopy was also used to measure the electron and hole dopants in graphene;[21,22] even the electronic structure of bilayer graphene was probed by resonant Raman scattering.[23] However, all the Raman studies above were carried out on micromechanical cleavage graphene (MCG). In this paper, we performed Raman studies of EG grown on SiC substrates. All the Raman peaks of EG were assigned and they differ substantially from that of MCG. Significant blue shifts of all the Raman peaks were observed which were attributed to the compressive strain caused by the SiC substrate. For thicker EG, the strain relaxes and the Raman peaks shift toward to those of MCG and graphite.

**Experimental**

The EG samples in this experiment were prepared by the following process: annealing a chemically etched (10% HF solution) n-type Si-terminated 6H-SiC (0001) sample (CREE Research Inc.) at 850 $^0$C under a silicon flux for 2 min in ultrahigh vacuum (UHV) resulting in a Si-rich 3x3-reconstructed surface, and subsequently annealing the sample several times at 1300 $^0$C in the absence of the silicon flux to form EG.[11,12,24] EG on C-terminated 6H-SiC(0001) was prepared in a similar way but in the absence of a silicon flux. The structure of EG was confirmed by in-situ Low-Energy-Electron- Diffraction (LEED), STM, and photoemission spectroscopy (PES).[25] The thickness of the EG layer was measured by monitoring the attenuation of the bulk SiC component in the Si 2p PES signal. The MCG was prepared by micromechanical cleavage and transferred to Si wafer with a 300 nm $SiO_2$ cap layer.[2] Phase contrast spectroscopy [25] was used to locate and determine the thickness of MCG. Raman spectra were recorded with a WITEC CRM200 Raman system. The excitation source was a 532 nm laser (2.33 eV) with power below 0.1 mW to avoid laser induced surface heating. The laser spot size is around 500 nm in diameter focused by a 100x optical lens (NA=0.95). The Raman spectra are recorded under the conditions and normalized in the figures to have the similar scale. The spectra resolution of our Raman system is ~1 cm$^{-1}$

**Results and discussion**

Figure 1 shows the LEED pattern (a) and a 5 nm$^2$ STM image (b) of EG grown on SiC substrate. In the LEED pattern, the pronounced spots of the (1×1) graphene lattice are clearly shown. Besides this, the SiC (1×1) pattern can also be observed. In Figure 1b, the dark spots reveal graphene (1×1) lattice. The six C atoms (as illustrated by small circles) surrounding each dark spot give the bright signal, which leads to a honeycomb atomic pattern. Therefore, both the LEED and STM reveal the graphene structure of our EG samples.

Since the characteristic STM images of carbon nanomesh and single layer graphene are quite different, the completion of single layer graphene can be determined by monitoring the phase evolution from carbon nanomesh to the single layer graphene by STM during the thermal annealing of SiC in UHV condition. In our experiments, the single layer graphene sample was obtained when the SiC surface was fully covered by graphene as checked by in-situ STM measurements.[17,26] However, the STM images for single layer and bilayer graphene on SiC are very similar. It is very hard to determine the layer thickness using this method. Instead, layer thickness for bilayer or thicker graphene sample is measured by monitoring the attenuation of the bulk SiC related Si 2p PES signal (photon beam energy is 500 eV) with normal emission condition. By using a simple attenuation model involving graphene layer on top of bulk SiC, the thickness of the graphene can be estimated using Equation (1) under normal emission condition: [24,27]

$$\frac{I_{SiC}^{Graphene}}{I_{SiC}^{Bulk}} = \exp(-t/\lambda) \quad (1)$$

where $I_{SiC}^{Graphene}$ is the normalized peak area intensity of Si 2p peak for graphene sample, $I_{SiC}^{Bulk}$ is the normalized peak area intensity of Si 2p peak for bulk SiC, $t$ is the thickness of the graphene layer. $\lambda$ is the electron escape depth in graphite (here we use the value of $\lambda$ in graphite instead of graphene). It can be obtained via the equations of $\lambda = a\lambda_m = 538E^{-2} + 0.41(aE)^{1/2}$, where $a$ is the layer thickness of graphite (0.355 nm), $E$ is the photon electron energy above Fermi level for Si 2p (~ 500 eV), and $\lambda_m$ electron attenuation length in monolayer.[28] $\lambda$ for electrons with kinetic energy of 500 eV (Si 2p photoelectrons) in graphite is calculated to be about 1.7 nm.

Figure 2 shows the Raman spectra of single and two-layer EG (grown on Si terminated SiC), MCG, bulk graphite, and SiC substrate. The 6H-SiC has several overtone peaks in the range of 1000 to 2000 cm$^{-1}$. The peak near ~1520 cm$^{-1}$ is the overtone of the TO(X) phonon at 761 cm$^{-1}$. The peak near ~1713 cm$^{-1}$ is a combination of optical phonons with wave vectors near the M point at the zone edges.[29,30] The weak SiC peak at ~1620 cm$^{-1}$ is not observable in our EG samples. The Raman spectrum of single layer EG has five peaks, located at 1368, 1520, 1597, 1713, and 2715 cm$^{-1}$, of which the peaks at 1520 and 1713 cm$^{-1}$ are from the SiC substrate. The 1368 cm$^{-1}$ peak is the so-called defect-induced D band; the 1597 cm$^{-1}$ peak is the in-plane vibrational G band; and the 2715 cm$^{-1}$ peak is the two-phonon 2D band.[31] The Raman signal of single layer MCG is much stronger (~10 times) than that of EG on SiC substrate. It is even comparable to that of bulk graphene. This phenomenon can be explained by the interference enhancement of Raman single of graphene on 300 nm SiO$_2$/Si substrate.[32] Compared with MCG and graphite, the Raman spectrum of EG shows the defect-induced D band, indicating that it contains defects, which may result from the surface dislocations, the corrugation, the interaction of graphene with substrate, or vacancies. The 2D band of single layer EG is broader than that of MCG, which is 60 cm$^{-1}$ compared to 30 cm$^{-1}$,[19,33] which can be explained by the poorer crystallinity of EG. However, compared to two- layer EG, the 2D band of single layer EG is still much narrower (60 cm$^{-1}$ compared to 95 cm$^{-1}$) and has a lower frequency (2715 cm$^{-1}$ compared to 2736 cm$^{-1}$), which are characteristics of single layer graphene. This has been widely used to identify single layer graphene of MCG.[19] Our Raman results confirm again that the EG on SiC is single and two layers, agree with the STM and PES identification. Another important observation was that the G (1597 cm$^{-1}$) and 2D (2715 cm$^{-1}$) bands of single layer EG shift significantly towards higher frequency from those of G (1580 cm$^{-1}$) and 2D (2673 cm$^{-1}$) of single layer MCG. Although the G band of single and few layer MCG may fluctuate ($\pm 3$ cm$^{-1}$) around the frequency of bulk graphite G band (1580 cm$^{-1}$), while the 2D bands of MCG may locate between 2660 and 2680 cm$^{-1}$,[20] the significant shifts of G band (17 cm$^{-1}$) and 2D band (42 cm$^{-1}$) of EG should be due to other mechanisms. The possibility that local electron/hole doping[21,22,34] in EG causes this Raman blueshift is not high, as it needs an electron/hole concentration more than $1.5 \times 10^{13}$ to induce the 17 cm$^{-1}$ blueshift of Raman G band.[35] It is shown that the dependence on doping of the shift of 2D-band is very weak and ~10-30% compared to that of G-band.[21,36] Therefore, the 42 cm$^{-1}$ 2D-band shift is too large to be achieved by electron/hole doping. Here, we attribute it to the interaction of SiC substrate with EG, most probably the strain effect, whereby the strain changes the lattice constant of graphene, hence the Raman peak frequencies.

To illustrate the origin of the strain, Figure 3 shows the schematic (top view (a) and side view (b)) of a graphene layer on SiC (0001) 6R3 reconstructed surface. The green, yellow, gray spheres represent C atoms in graphene, Si atoms in SiC, and C atoms in SiC, respectively. The large black circles represent the 6R3 lattice. The bulk lattice constant we used for SiC is 3.073 Å,[37] while that for graphene is 2.456 Å.[38] It is obvious that the 13 × 13 graphene (31.923 Å) matches the 6R3 lattice (31.935 Å) quite well. On the other hand, the 2 × 2 graphene (4.9 Å) does not match the R3 structure (5.34 Å) (small black circles). Our previous STM results showed that the 6R3 surface did not always retain its "6 × 6" periodicity. The pore size of honeycombs in STM can be changed from 20 Å to 30 Å, depend on the annealing temperature.[17] Hence, this surface can be described as a dynamic superstructure formed by the self-organization of surface carbon atoms at high temperatures. That is the reason we prefer to denote it as carbon nanomesh instead of 6R3. As a result, the mismatch between graphene 13 × 13 lattice (~32 Å) and carbon nanomesh (20 to 30 Å) will cause the compressive strain on EG. Calizo et al. studied the substrate effect of MCG and they did not observe such a strong stress effect,[39] partially because the weak interaction between MCG and substrates (Van der waals force) is not strong enough.

Graphene has a very thin (2D structure) and its stress induced by the lattice mismatch with the SiC substrate can be considered as biaxial. The biaxial compressive stress on EG can be estimated from the shift of Raman $E_{2g}$ phonon (G band) with the following analysis.

For a hexagonal system, the strain ε induced by an arbitrary stress σ can be expressed as:[40,41]

$$\begin{bmatrix} \varepsilon_{xx} \\ \varepsilon_{yy} \\ \varepsilon_{zz} \\ \varepsilon_{yz} \\ \varepsilon_{zx} \\ \varepsilon_{xy} \end{bmatrix} = \begin{bmatrix} S_{11} & S_{12} & S_{13} & & & \\ S_{12} & S_{11} & S_{13} & & & \\ S_{13} & S_{13} & S_{33} & & & \\ & & & S_{44} & & \\ & & & & S_{44} & \\ & & & & & 2(S_{11}-S_{12}) \end{bmatrix} \begin{bmatrix} \sigma_{xx} \\ \sigma_{yy} \\ \sigma_{zz} \\ \sigma_{yz} \\ \sigma_{zx} \\ \sigma_{xy} \end{bmatrix}, \quad (2)$$

with the coordinate $x$ and $y$ in the graphite/graphene plane and $z$ perpendicular to the plane. In the case of biaxial stress:

$$\sigma_{xx} = \sigma_{yy} = \sigma \quad (3)$$

$$\sigma_{zz} = \sigma_{yz} = \sigma_{zx} = \sigma_{xy} = 0 \quad (4)$$

So that:

$$\varepsilon_{xx} = \varepsilon_{yy} = (S_{11} + S_{12})\sigma \quad (5)$$

$$\varepsilon_{zz} = 2S_{13}\sigma \quad (6)$$

$$\varepsilon_{yz} = \varepsilon_{zx} = \varepsilon_{xy} = 0 \quad (7)$$

The secular equation of such system is:

$$\begin{vmatrix} A(\varepsilon_{xx} + \varepsilon_{yy}) - \lambda & B(\varepsilon_{xx} - \varepsilon_{yy} + 2i\varepsilon_{xy}) \\ B(\varepsilon_{xx} - \varepsilon_{yy} + 2i\varepsilon_{xy}) & A(\varepsilon_{xx} + \varepsilon_{yy}) - \lambda \end{vmatrix} = 0 \quad, \quad (8)$$

where

$$\lambda = \omega_\sigma^2 - \omega_0^2 \tag{9}$$

with $\omega_\sigma$ and $\omega_0$ the frequencies of Raman $E_{2g2}$ phonon under stressed and unstressed conditions.

With all the shear components of strain equal to zero, equation (9) reduces to:

$$\begin{vmatrix} A(\varepsilon_{xx}+\varepsilon_{yy})-\lambda & B(\varepsilon_{xx}-\varepsilon_{yy}) \\ B(\varepsilon_{xx}-\varepsilon_{yy}) & A(\varepsilon_{xx}+\varepsilon_{yy})-\lambda \end{vmatrix} = 0 \tag{10}$$

There is only one solution for it:

$$\lambda = A(\varepsilon_{xx}+\varepsilon_{yy}) = 2A\varepsilon_{xx} = 2A(S_{11}+S_{12})\sigma \tag{11}$$

Therefore,

$$\omega_\sigma - \omega_0 = \frac{\lambda}{\omega_\tau + \omega_0} \approx \frac{\lambda}{2\omega_0} = \frac{A(S_{11}+S_{12})\sigma}{\omega_0} = \alpha\sigma \tag{12}$$

where $\alpha = \dfrac{A(S_{11}+S_{12})}{\omega_0}$ is the stress coefficient for Raman shift.

Using A= -1.44 × 10$^7$ cm$^{-2}$ [40] and graphite elastic constants $S_{11}$=0.98 × 10$^{-12}$ Pa$^{-1}$ and $S_{12}$= -0.16 × 10$^{-12}$ Pa$^{-1}$,[42] and $\omega_0$ =1580 cm$^{-1}$, the stress coefficient $\alpha$ is about 7.47 cm$^{-1}$/GPa. Hence, a biaxial stress of 2.27 GPa on EG is obtained from the 17 cm$^{-1}$ shift of G band frequency of EG compared to that of bulk graphite or MCG. The strong compressive stress may affect the properties of graphene (both physical and electronic properties), since strain/stress studies in CNTs have already shown many such examples.[43-45]

The Raman spectra of EG grown on Si terminated SiC (Si-SiC) and C terminated SiC (C-SiC) also show differences, as shown in Figure 4. Both samples were grown under similar conditions and are two layers in thickness. The EG on C-SiC has a broader G band, which means its crystallinity is worse than EG grown on Si-SiC.[46] Besides, it contains more defects demonstrated by a stronger defected-induced D band. The G bands of EG on C-SiC and Si-SiC have similar frequency (~1597 $cm^{-1}$), indicating that both EGs are affected by the substrates, and they are under similar stress. Interestingly, EG on C-SiC has much lower D and 2D band frequencies, which are at 1343 and 2682 $cm^{-1}$ compared to 1369 and 2736 $cm^{-1}$ for EG on Si-SiC substrate. As the G band frequencies of C-EG and Si-EG are similar, the difference in D and 2D band frequencies is not caused by strain. According to the double resonance theory, the Raman frequencies of the D and 2D bands show strong dependence on the electronic band structure as well as the excitation laser energy (fixed at 532nm in our experiments).[47] Hereby, we attribute the observation to the difference in the electronic structure of the two systems. Recently, calculations by Mattausch et al. also showed that the band structures of EG grown on Si-SiC and EG on C-SiC differ substantially.[14]

To investigate the evolution of thicker EG on SiC substrate, we grew EGs with different thickness and typical Raman spectra of EGs on C-SiC are shown in Figure 5. As the EG thicknesses increase, the Raman peaks (D, G and 2D) of EG shift to lower frequencies, towards that of bulk graphite. This can be easily understood since when the EG thickness increases, the effect of substrate on EG becomes weaker and the EG lattice relaxes.

**Conclusion**

In summary, Raman spectroscopic studies of epitaxial graphene grown on SiC substrates were carried out. All the Raman peaks of EG have been assigned and compared with those of MCG and bulk graphite. The results show that graphene grown on SiC is compressive stressed. The lattice mismatch between 13 × 13 graphene and carbon nanomesh is used to explain the origin of stress. Using a biaxial stress model, the compressive stress on EG was estimated to be about 2.27 GPa, which affects the optical and electronic properties of graphene similar to what has been observed in CNTs. Finally, from the Raman spectra difference of EG on Si-SiC and C-SiC, we demonstrate that the electronic band structure of EG grown on Si-SiC and C-SiC are quite different. Our findings should provide useful information for understanding the interaction between EG and substrate as well as the potential device applications of EG-based nanodevices.

# References

[1] A. K. Geim, K. S. Novoselov, Nature Materials **6**, 183 (2007).

[2] K. S. Novoselov, A. K. Geim, S. V. Morozov, D. Jiang, Y. Zhang, S. V. Dubonos, I. V. Grigorieva, A. A. Firsov, Science **306**, 666 (2004).

[3] K. S. Novoselov, A. K. Geim, S. V. Morozov, D. Jiang, M. I. Katsnelson, I. V. Grigorieva, S. V. Dubonos, A. A. Firsov, Nature **438**, 197 (2005).

[4] G. W. Semenoff, Phys. Rev. Lett. **53**, 2449 (1984).

[5] F. D. M. Haldane, Phys. Rev. Lett. **61**, 2015 (1988).

[6] Y. B. Zhang, Y. W. Tan, H. L. Stormer, P. Kim, Nature **438**, 201 (2005).

[7] J. Scott Bunch, Arend M. van der Zande, Scott S. Verbridge, Ian W. Frank, David M. Tanenbaum, Jeevak M. Parpia, Harold G. Craighead, Paul L. McEuen, Science **315**, 490 (2007).

[8] Hubert B. Heersche, Pablo Jarillo-Herrero, Jeroen B. Oostinga, Lieven M. K. Vandersypen, Alberto F. Morpurgo, Nature **446**, 56 (2007).

[9] Jannik C. Meyer, A. K. Geim, M. I. Katsnelson, K. S. Novoselov, T. J. Booth, S. Roth, Nature **446**, 60 (2007).

[10] T. Ohta, A. Bostwick, T. Seyller, K. Horn, E. Rotenberg, Science **313**, 951 (2006).

[11] C. Berger, Z. M. Song, X. B. Li, X. S. Wu, N. Brown, C. Naud, D. Mayo, T. B. Li, J. Hass, A. N. Marchenkov, E. H. Conrad, P. N. First, W. A. de Heer, Science **312**, 1191 (2006).

[12] C. Berger, Z. M. Song, X. B. Li, A. Y. Ogbazghi, R. Feng, Z. T. Dai, A. N. Marchenkov, E. H. Conrad, P. N. First, W. A. de Heer, J. Phys. Chem. B **108**, 19912 (2004).

[13] S. Y. Zhou, G. H. Gweon, A. V. Fedorov, P. N. First, W. A. De Heer, D. H. Lee, F. Guinea, A. H. Castro Neto, A. Lanzara, Nature Materials **6**, 770 (2007).

[14] A. Mattausch, O. Pankratov, Phys. Rev. Lett. **99**, 076802 (2007).

[15] F. Varchon, R. Feng, J. Hass, X. Li, B. N. Nguyen, C. Naud, P. Mallet, J. Y. Veuillen, C. Berger, E. H. Conrad, L. Magaud Phys. Rev. Lett. 99, 126805 (2007).

[16] J. Hass, R. Feng, J. E. Millan-Otoya, X. Li, M. Sprinkle, P. N. First, W. A. De Heer, E. H. Conrad, Phys. Rev. B **75**, 214109 (2007).

[17] W. Chen, H. Xu, L. Liu, X. Y. Gao, D. C. Qi, G. W. Peng, S. C. Tan, Y. P. Feng, K. P. Loh, A. T. S. Wee, Surf. Sci. **596**, 176 (2005).

[18] P. Mallet, F. Varchon, C. Naud, L. Magaud, C. Berger, J. Y. Veuillen, Phys. Rev. B **76**, 041403(R) (2007).

[19] A. C. Ferrari, J. C. Meyer, V. Scardaci, C. Casiraghi, M. Lazzeri, F. Mauri, S. Piscanec, D. Jiang, K. S. Novoselov, S. Roth, A. K. Geim, Phys. Rev. Lett. **97**, 187401 (2006).

[20] D. Graf, F. Molitor, K. Ensslin, C. Stampfer, A. Jungen, C. Hierold, L. Wirtz, Nano Lett. **7**, 238 (2007).

[21] J. Yan, Y. B. Zhang, P. Kim, A. Pinczuk, Phys. Rev. Lett. **98**, 166802 (2007).

[22] S. Pisana, M. Lazzeri, C. Casiraghi, K. S. Novoselov, A. K. Geim, A. C. Ferrari, F. Mauri, Nature Materials **6**, 198 (2007).

[23] L. M. Malard, J. Nilsson, D. C. Elias, J. C. Brant, F. Plentz, E. S. Alves, A. H.

Castro Neto, M. A. Pimenta Preprint at http://arxiv.org/abs/ cond-mat / 0708.1345 (2007).

[24] W. Chen, S. Chen, D. C. Qi, X. Y. Gao, A. T. S. Wee, J. Am. Chem. Soc. **129**, 10418 (2007).

[25] Z. H. Ni, H. M. Wang, J. Kasim, H. M. Fan, T. Yu, Y. H. Wu, Y. P. Feng, Z. X. Shen, Nano Lett. **07**, 2758 (2007).

[26] C. Riedl, U. Starke, J. Bernhardt, M. Franke, K. Heinz Phys. Rev. B **76**, 245406 (2007)

[27] Charles S. Fadley Prog. Surf. Sci. **16**, 275 (1984)

[28] M. P. Seah, W. A. Dench, Surf. Interface Anal. **1**, 2 (1979)

[29] J. C. Burton, L. Sun, F. H. Long, Z. C. Feng, I. T. Ferguson Phys. Rev. B **59**, 7282 (1999).

[30] H. W. Kunert, T. Maurice, J. Barnas, J. Malherbe, D. J. Brink, L. Prinsloo, Vacuum **78**, 503 (2005).

[31] E. B. Barros, N. S. Demir, A. G. Souza Filho, J. Mendes Filho, A. Jorio, G. Dresselhaus, M. S. Dresselhaus, Physical Review B **71**, 165422 (2005).

[32] Y. Y. Wang, Z. H. Ni, H. M. Wang, Y. H. Wu, Z. X. Shen, Applied Physics Letters 92, 043121 (2008)

[33] I. Calizo, A. A. Balandin, W. Bao, F. Miao, C. N. Lau, Nano Letters, **7**, 2645 (2007)

[34] C. Casiraghi, S. Pisana, K. S. Novoselov, A. K. Geim, A. C. Ferrari, Appl. Phys. Lett. **91**, 233108 (2007).

[35] A. Das, S. Pisana, S. Piscanec, B. Chakraborty, S. K. Saha, U. V. Waghmare, R. Yiang, H. R. Krishnamurhthy, A. K. Geim, A. C. Ferrari, A. K. Sood, Preprint at http://arxiv.org/abs/ 0709.1174v1 (2007)

[36] C. Stampfer, F. Molitor, D. Graf, K. Ensslin, A. Jungen, C. Hierold, and L. Wirtz Appl. Phys. Lett. **91**, 241907 (2007)

[37] L. S. Ramsdell, American Mineralogist **29**, 519 (1944).

[38] J. S. Kukesh, L. Pauling, American Mineralogist **35**, 125 (1950).

[39] I. Calizo, W. Z. Bao, F. Miao, C. N. Lau, A. A. Balandin, Appl. Phys. Lett. **91**, 201904 (2007).

[40] H. Sakata, G. Dresselhaus, M. S. Dresselhaus, M. Endo J. Appl. Phys. **63**, 15 (1988).

[41] J. W. Ager III, S. Anders, A. Anders, I. G. Brown, Appl. Phys. Lett. **66**, 19, (1995).

[42] M. S. Dresselhaus, G. Dresselhaus, K. Sugihara, I. L. Spain, H. A. Goldberg, Graphite Fibers and Filaments (Springer, Heidelberg, 1988).

[43] E. D. Minot, Y. Yaish, V. Sazonova, J. Y. Park, M. Brink, P. L. McEuen, Phys. Rev. Lett. **90**, 156401 (2003).

[44] T. W. Tombler, C. Zhou, L. Alexseyev, J. Kong, H. Dai, L. Liu, C. S. Jayanthi, M. Tang, S. Wu, Nature **405**, 769 (2000).

[45] R. Heyd, A. Charlier, E. McRae, Phys. Rev. B **55**, 6820 (1997).

[46] J. M. Holden, P. Zhou, X. X. Bi, P. C. Eklund, S. Bandow, R. A. Jishi, K. Das Chowdhury, G. Dresselhaus, M. S. Dresselhaus, Chem. Phys. Lett. **220**, 186 (1994).

[47] C. Thomsen, S. Reich, Phys. Rev. Lett. **85**, 5214 (2000).

Figure captions:

Figure 1. (Color online) (a) LEED pattern of epitaxial graphene on 6H-SiC(0001). Incident beam energy: 175 eV. (b) 5 × 5 nm$^2$ STM image of epitaxial graphene on 6H-SiC(0001).

Figure 2. Raman spectra of single and two-layer EG grown on SiC, SiC substrate, MCG, and bulk graphite as indicated. The inset is an enlarged part of the 2D band region of single and two-layer EG. The hollow symbols are experimental data and the solid line is the fitted curve.

Figure 3. (Color online) Schematic (top view (a) and side view (b)) of a graphene layer on SiC (0001) surface. The green, yellow and gray spheres represent C in graphene, Si in SiC and C in SiC, respectively. The SiC surface was after 6R3 reconstruction and a 13×13 graphene lattice lies on above it. The small black circles represent the R3 lattice, while the large black circles represent the 6R3 lattice.

Figure 4. Raman spectra of EG grown on Si terminated SiC (Si-SiC) and C terminated SiC (C-SiC)

Figure 5. Raman spectra of EGs on C-SiC substrate of different thickness.

Figure 1.

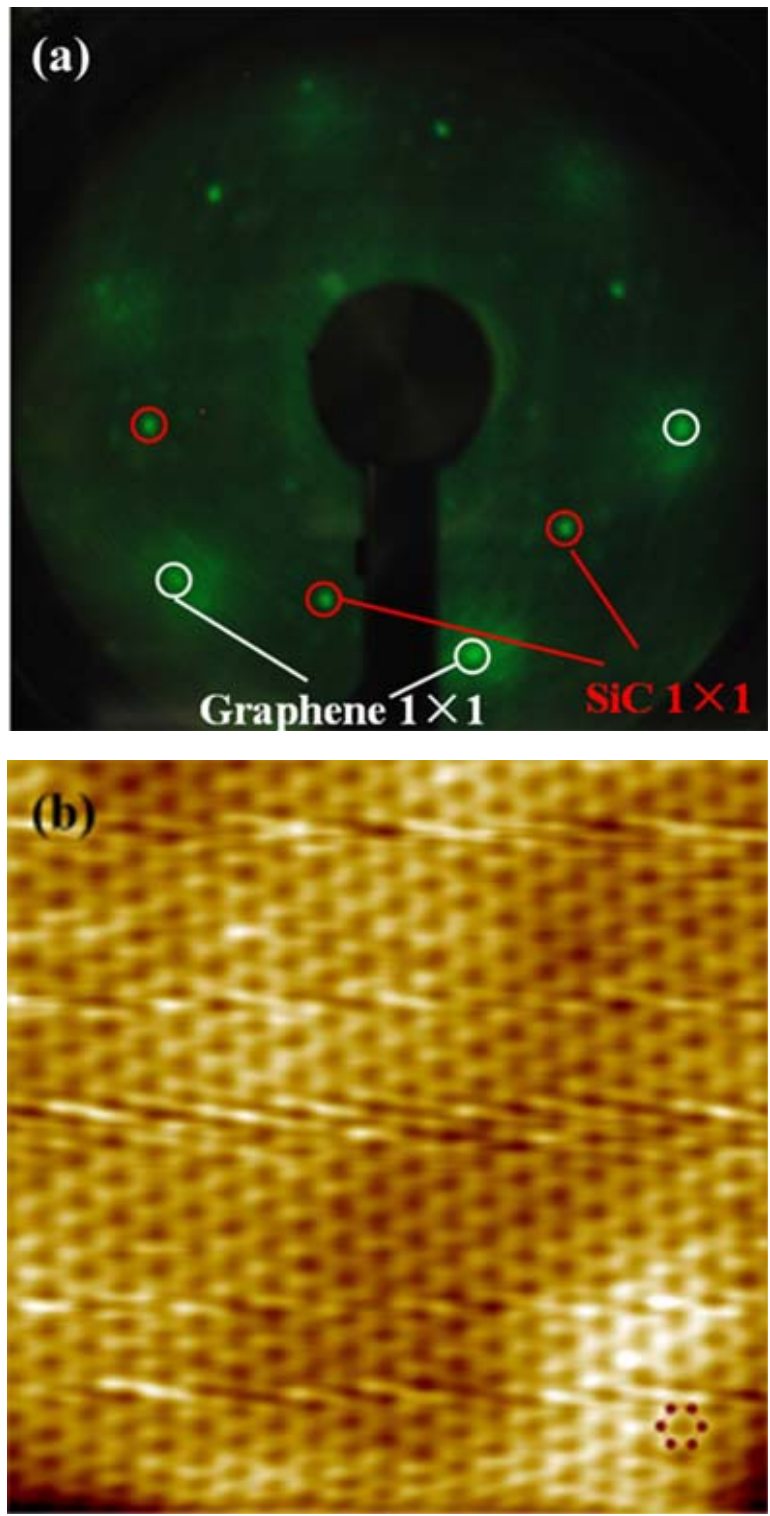

Figure 1. (Color online) (a) LEED pattern of epitaxial graphene on 6H-SiC(0001). Incident beam energy: 175 eV. (b) 5 × 5 nm$^2$ STM image of epitaxial graphene on 6H-SiC(0001).

Figure 2.

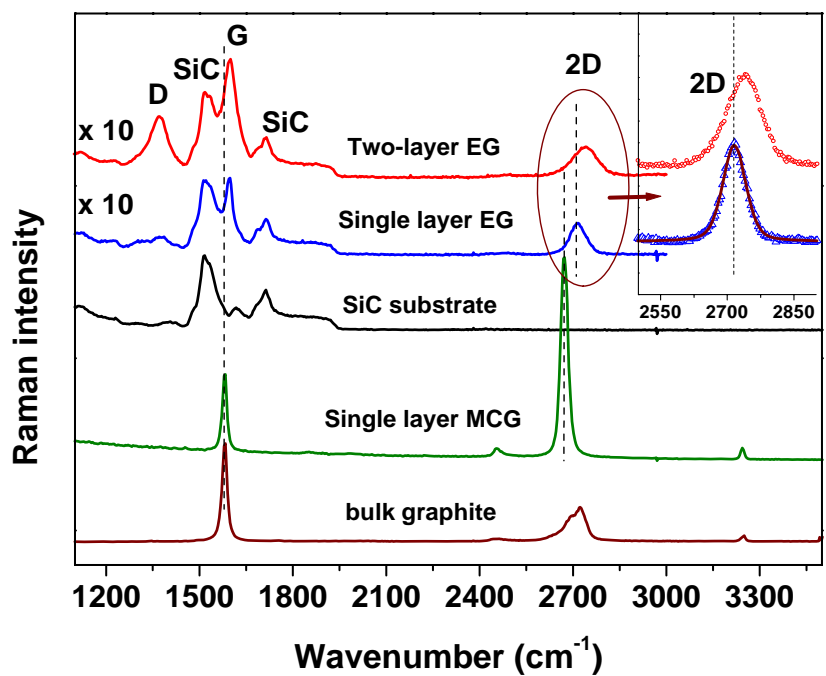

Figure 2. Raman spectra of single and two-layer EG grown on SiC, SiC substrate, MCG, and bulk graphite as indicated. The inset is an enlarged part of the 2D band region of single and two-layer EG. The hollow symbols are experimental data and the solid line is the fitted curve.

Figure 3.

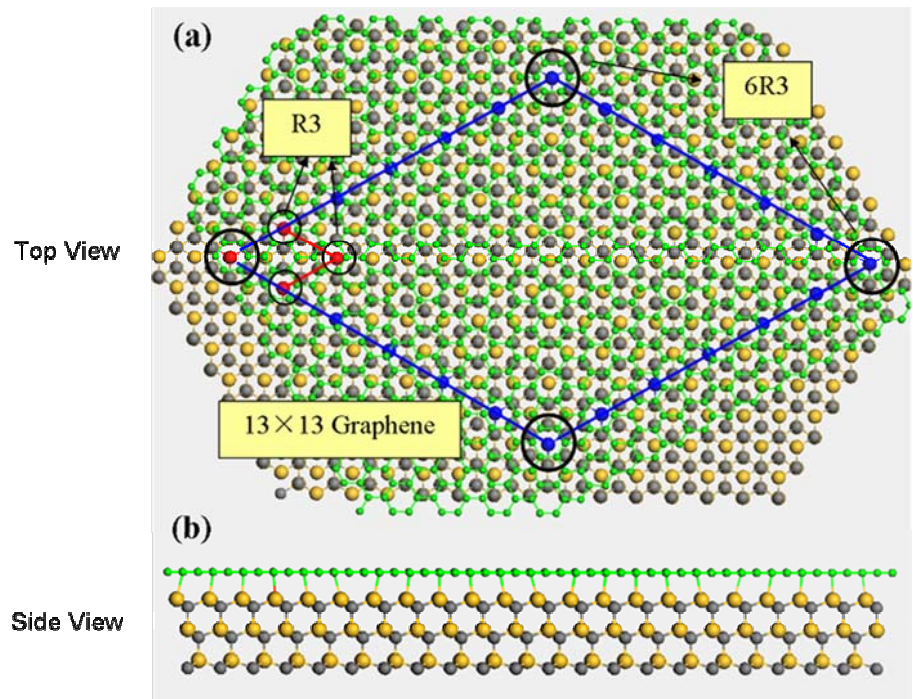

Figure 3. (Color online) Schematic (top view (a) and side view (b)) of a graphene layer on SiC (0001) surface. The green, yellow and gray spheres represent C in graphene, Si in SiC and C in SiC, respectively. The SiC surface was after 6R3 reconstruction and a 13×13 graphene lattice lies on above it. The small black circles represent the R3 lattice, while the large black circles represent the 6R3 lattice.

Figure 4.

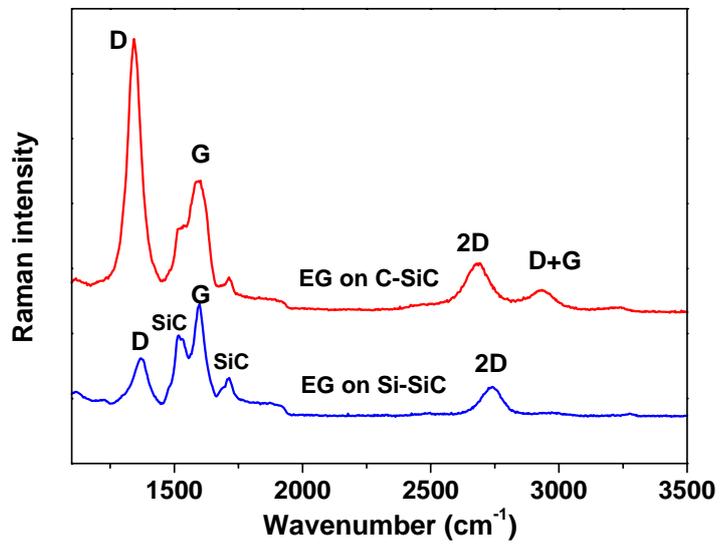

Figure 4. Raman spectra of EG grown on Si terminated SiC (Si-SiC) and C terminated SiC (C-SiC).

Figure 5.

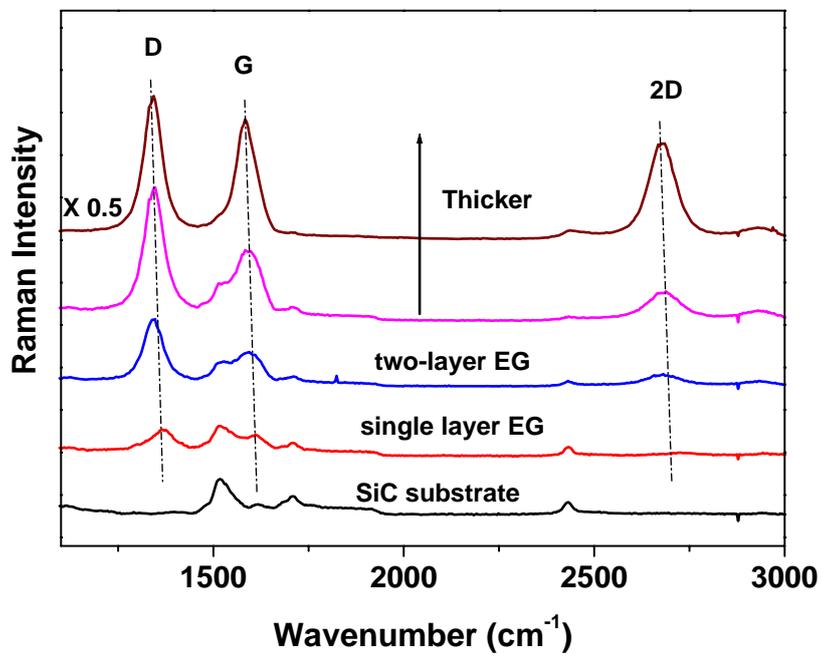

Figure 5. Raman spectra of EGs on C-SiC substrate of different thickness.